\documentstyle[11pt,epsf]{article}
\newcommand{\be}{\begin{equation}}
\newcommand{\ee}{\end{equation}}
\newcommand{\bear}{\begin{eqnarray}}
\newcommand{\ear}{\end{eqnarray}}
\newcommand{\no}{\noindent}
\date{}
\newcommand{\e}{\mbox{e}}

\newcommand{{\Tr}}{\rm Tr}
\newcommand{{\tr}}{\rm tr}
\newcommand{\slD}{\raise.15ex\hbox{$/$}\kern-.57em\hbox{$D$}}

\newcommand{\tow}[2]{{\!\begin{array}{c}
{#1}\vspace*{-0.8cm}\\
{#2}\vspace*{-0.1cm}\end{array}\!}}

\def\A{\bf A}

\begin{document}

\begin{flushright} HD--THEP--01-47\end{flushright}
\begin{center}
\vspace{.5cm}
{\bf\LARGE
Determinant Calculations }\\
\vspace{.3cm}
{\bf\LARGE  with Random Walk Worldline Loops}\\
\vspace{1cm}
Michael G. Schmidt$^{a\,}$\footnote{m.g.schmidt@thphys.uni-heidelberg.de}
and Ion-Olimpiu Stamatescu$^{a,b\,}$\footnote{stamates@thphys.uni-heidelberg.de}\\
\vspace{1cm} {\em
$^a$Institut f\"ur Theoretische Physik, Philosophenweg 16,
D-69120 Heidelberg, Germany\\[.1cm]
$^b$Forsch. der Ev. Studiengemeinschaft, Schmeilweg 5, D-69118, Heidelberg, Germany
}
\end{center}

\vskip2cm
\begin{abstract}
We propose to calculate bosonic and fermionic determinants
with some general field background, and the
corresponding 1-loop effective actions  by evaluating random walk  worldline
loops generated statistically on the lattice. This is illustrated by some numerical calculations
for simple gauge field backgrounds and then discussed for the general
case.
\end{abstract}
\newpage
\section{Introduction}

The discussion of bosonic and fermionic determinants in some background
and of the related 1-loop effective action is most transparent in the
worldline/relativistic particle formalism \cite{1}-\cite{3}. The Euler-Heisenberg 
effective action of a constant electromagnetic field \cite{4},
for instance, can be presented
very elegantly in this way \cite{3}.
Also for general gauge field backgrounds this
formulation is  very helpful allowing for an economical inverse
mass (derivative) expansion to high order \cite{5}.
For background fields like an instanton,
sphaleron or a bounce solution used in tunneling problems the evaluation
of a derivative expansion \cite{6} and alternatively the approach
based on solutions for the eigenvalue spectrum \cite{7}
are rather sophisticated. The nonabelian gauge
field groundstate background in 3- and 4-dimensional QCD and in
the hot electroweak theory is even less accessible to an analytical
treatment. For instance, the discussion of a nonperturbative magnetic and
tachyonic mass in 3- and 4-dimensional SU(2) theory
is very difficult even if one uses simplified assumptions
about the gauge field vacuum \cite{8}. Finally, we should notice
that 2-loop effective actions are also accessible \cite{9,3,10}

In this paper we consider the
numerical evaluation of quantum field theoretical determinants
in the worldline formalism using  lattice methods.
We consider the bosonic determinant in
a gauge field background
\bear
\Gamma(A)&=& -\log\det ({\cal D}^2 + m^2)
          =   -\tr\log ({\cal D}^2 + m^2)   \nonumber \\
&=&\int^\infty_0\frac{dT}{T}\int d^Dx\int _{x(0)=x(T)=x}
[{ D}x]\,\tr_P\exp\left\{-\int^T_0d\tau\left(\frac{\dot x^2}{4}+
ig{\bf A}_\mu\dot x_\mu+m^2
\right)\right\} \nonumber\\
&=&\int
d^Dx\int^\infty_0\frac{dT}{T}e^{-m^2T} \int _{x(0)=x(T)=x}
[Dx]\,
\exp\left\{-\int^T_0 d\tau\frac{\dot x^2}{4}\right\}
\label{e.wlac} \\
&&\times \, \tr_P\exp
\left\{-ig\oint {\A}_\mu dx_\mu\right\}, \nonumber
\ear
where the worldline sum (here already in the Euclidean formulation
adequate for the lattice analysis) is
over all closed paths and $\tr_P$ indicates path ordering.
In proceeding to the lattice
formulation of the problem it is natural to consider the loops on the same
(usually hypercubic) lattice, on which the general gauge field configurations
are generated. This appears to be difficult if one wants to enforce the
$\exp\{-\int^T_0 d\tau\frac{\dot x^2}{4}\}$ weight for randomly chosen paths,
 as proposed in an interesting recent paper \cite{11}. However,
as is well known,  random walk paths automatically implement  the above
integral weight
of the free particle propagation on a worldline \cite{12}.
The expression (\ref{e.wlac})
can thus be discretized on a D-dimensional lattice $\Lambda_D$ as
\be
\label{e.wlal}
\Gamma^{discr}(A) =\sum_{x\in \Lambda_D} {\cal L}(A,x);\ \
{\cal L}(A,x) = \sum_L\frac{1}{L}e^{-m^2 \frac{L}{2D}}
\left(\frac{1}{2D}\right)^L \sum_{\{\omega_L(x)\}}
\tr \left( \prod_{l \in \omega_L(x)} U_l \right),
\ee
where dimensionfull quantities are understood as given in units of the lattice
scale $a$ according to their naive dimension
and we have also discretized $T$ using the $\tau$
intervals
\be
\Delta \tau=\frac{T}{L}=\frac{a^2}{2D}
\label{e.dtau}
\ee
Here $\{\omega_L(x)\}$ is the set of all closed lattice paths with length
$|\omega_L|=L$  obtained by random walk (RW) starting and ending at $x$,
and we have written down
explicitly the random walk ``measure" $(2D)^{-L}$
(to be implemented by the
actual RW procedure). The mass dependence
$e^{-m^2T}$ could be also obtained by  a random walk weight but we prefer
to keep this dependence explicit. $U_l$ are link variables.

    On the lattice the expansion according to the loop lengths (\ref{e.wlal})
is of course equivalent to the
   usual hopping parameter expansion for the logarithm of the determinant, 
both for bosons
and for fermions (see, e.g., \cite{12}, \cite{kawa},
\cite{stadet}). This can be seen explicitly by rewriting (e.g., 
in the bosonic case)
\be
\kappa^L =(m^2+ 2 D)^{-L} \tow{\simeq}{m^2 \ll 2D} (2D)^{-L} \e^{- 
m^2 \frac{L}{2D}}
\ee
(and similarly for fermions).  The number of different loops of length $L$ increases exponentially with 
the length,
therefore one usually must cut the hopping parameter
expansion at a rather small $L$. We suggest here to use
a statistical procedure for sampling loops in a wide interval of lengths,
instead of summing over all possible 
loops up to a given order
(see also \cite{phil}). In our 
approach we use the random walk representation (\ref{e.wlal}) to construct a 
{\it (sub-)ensemble} of loops of various lengths and evaluate the determinant
with help of this sub-ensemble. By construction, the loops appear with the 
correct probabilities, and since this is a statistical procedure  
 we  may expect to have only statistical errors. Since 
loops of length $L$ are typically relevant
 at the scale $\sqrt{L}$, arbitrarily cutting the 
loop expansion at some limited $L$ may completely miss  physical effects at 
scales larger than  $\sqrt{L}$. In our method we trade these systematic
errors for statistical ones such that we may hope to reproduce also such effects,
albeit within statistical uncertainties. Notice that, on the one hand,
we cannot use too rough lattices since this 
introduces discretization errors. On the other hand, the contribution of large loops is
 damped even at small mass by a 
factor $\propto L^{-D/2}$, see (\ref{e.nrl}) next section. This suggests that,
depending on the effects we want to observe, there is an optimal range of
loops to be taken into account. In 
practice we use loops of length up to $L \simeq 200$ on lattices of $32^D$.
For the cases studied here these ensembles appeared adequate.

Chapter 2 describes the procedure how to generate closed loops
and checks that the case without interaction on our loop ensembles
 converges to the well-known
continuum (large $L$) result.
In expression (\ref{e.wlal}) we can insert some (quasi) classical
background gauge field $A_\mu$ discretized by lattice connections
$U$. This we will do for a constant magnetic field in chapter 3.
This case allows comparison with the well-known Euler-Heisenberg
action in its worldline formulation. We also treat there the case
of constant $SU(2)$ fields in two perpendicular 2-planes in
four dimensions which has a topological meaning \cite{13}. The case of a
constant magnetic field in a half space treated in ref. \cite{11},
which introduces a correlation scale (penetration length)
is also studied.

For a quantum gauge field vacuum the action
$\Gamma^{discr}(A)$ has to be inserted into a gauge field path
integral discretized on the lattice to obtain the (quenched) vacuum energy
\be\label{e.qcd1}
V(\beta) = \langle \Gamma \rangle = 
\sum_L\frac{1}{L}e^{-m^2 \frac{L}{2D}}
\left(\frac{1}{2D}\right)^L \sum_{\{\omega_L
\}}\frac{1}{Z_{\rm YM}}\int [DU]\e^{-\beta S_{\rm YM}(\{U\})}
\tr\left( \prod_{l \in \omega_L} U_l \right)
\ee
with the Yang-Mills discretized action $S_{\rm YM}$ and partition function $Z_{\rm YM}$. In the case of a further
field $\phi$ in the background coupling to the loop a term $V''(\phi(x))$
has to be introduced in the discretized worldline action and further
weights the sum over random paths. These 1-loop actions in a general
background are  our real goal to be
accomplished in future work.
Chapter 4 presents a final discussion and an outlook to the case of
spin in the loop, in particular to fermionic determinants.

\section{The lattice loop ensemble}

As remarked in  the introduction we use random walk to
produce a loop ensemble thermalized with the ``Boltzmann factor"
\be
\e^{- \frac{1}{4}\sum_{i=0}^{L-1}
\sum_{\mu=1}^D (n_{\mu}(i+1) - n_{\mu}(i))^2 |_{n_{\mu}(0)=n_{\mu}(L)}},
\label{e.lalb}
\ee
see (\ref{e.wlal}), which
is the discretized form of the corresponding factor in
(\ref{e.wlac}). In reconstructing the whole
path integral we only need to weight the RW loop
contributions $\tr\left( \prod_{l \in \omega_L} U_l \right)$ with
\be
L^{-1} \e^{-m^2 \frac{L}{2D}}
\ee
since the factor $(2D)^{-L}$ is implicit in the RW generation. From now on
we denote by $\{\omega\}$, $\{\omega_L\}$ the corresponding ensemble
 of loops {\it as it is produced} by the
random walk procedure, that is,  with the frequencies with which the loops
are generated in the actual procedure.

In generating the loops we specify a maximal number of trials ${\cal N}$  (e.g., 2000000
for the $32^3$ lattice) for each given number of steps (between 4 and 180). We
first generate an $\{\omega(x_\mu = 1)\}$ ensemble of loops by starting 
 the random walk at the point
 $x_{\mu} = 1$ of
the periodic lattice and collecting the loops which after the prefixed
number of steps
have returned to $x_{\mu} = 1$, independently on possible self-crossings or
retracing. We define ${\cal L}(A,1)$ using this $\{\omega(x_\mu = 1)\}$
ensemble (for the application in sect. \ref{s.mghs} we first shift the 
approximate ``center of mass" of the
loops to $x_\mu=1$). For any given $x$ we can calculate ${\cal L}(A,x)$ 
in (\ref{e.wlal})
 by accordingly shifting the loops of the $\{\omega(x_\mu = 1)\}$
ensemble as a whole. On the other hand, the full loop
ensemble, that is, the $x$-summation $\sum_{x\in \Lambda_D}$ in
(\ref{e.wlal}), is approximated by
considering  sufficiently many random translations of each of the
loops of the  $\{\omega(x_\mu = 1)\}$ ensemble (for
homogeneous field configurations this is, of course, redundant).
That is, loop and $x$ ``summations" are done simultaneously.
This construction, i.e. generation of loops at $x=1$ and 
stochastic summation over  $x$ by random translations,
automatically reproduces (statistically) the correct 
loop multiplicities in the hopping parameter expansion, 
by which a certain loop of length $L$ appears $L$ times, unless it represents
$r$ windings of a loop of length $L'$, $L=r\,L'$, in which 
case it only appears  $L'$ times \cite{12}, \cite{stadet}.
To improve the loop ensemble we can
implement random rotations of the loops (before translations). Since the
loop evaluation is vectorized over rotations and translations, these are
reasonably cheap.
In the
following we work in $D=3$ and $D=4$ and use  an $\{\omega(x_\mu = 1)\}$
ensemble of up to 600000 (250000) loops on lattices of $N_x\times N_y
\times N_z (\times N_t)$ with $N_{\mu}\, = \,12,\, 16,\, 24$ and 32, and
periodic boundary conditions.

\begin{figure}[htb]
\vspace{9cm}
\includegraphics{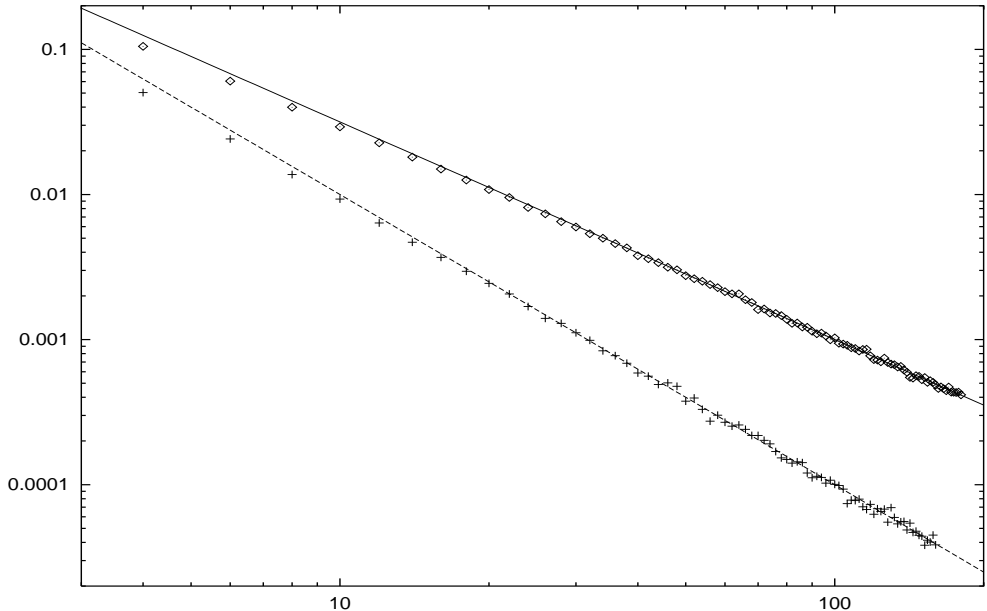}
\caption{{\it Number of closed loops of length $L$, $n_L$ generated
by random walk. 
We show the frequencies $\nu_L$ (\ref{e.freq}) vs $L$ 
in $D=3$ (diamonds)
and $D=4$ dimensions (crosses). 
The full lines represent the corresponding continuum results $L^{-D/2}$.
}
}
\label{f.free}
\end{figure}

For a free theory ($U_l = 1$ above) the contribution of the RW-loops
of length $L$ is proportional to their number, $n_L$ (which counts
the different loops with their multiplicities). From
(\ref{e.wlac}) we have:
\be
n_L \simeq T^{-D/2} \propto L^{-D/2}
\label{e.nrl}
\ee
for large $L$
-- this  well known relation can  be seen by choosing $A=0$, $m=0$ in
(\ref{e.wlac}) and rescaling $\tau
\rightarrow \tau T,\, x_{\mu} \rightarrow x_{\mu} \sqrt{T}$.
In Fig. \ref{f.free} we show the $n_L$ distribution
 in $D=3$ and 4 for our working 
ensembles of loops. We observe the excellent agreement with (\ref{e.nrl})
 for large $L$ and the expected
 systematic deviations for small $L$.
In the following we shall use the actually realized $n_L$ 
(instead of $L^{-D/2}$) to
normalize the loop contributions when necessary,
 to improve on the systematic errors. 

Notice that the number of
loops at large $L$ is much smaller than at small $L$. If large loops are
physically important this fact may lead to large statistical errors. One 
improves on this by using a larger number of trials at high $L$, producing 
in this way more loops there. Then we must renormalize their
contributions according to the number of trials. The loop frequencies are thus:
\be
\nu_L = \frac{2\,n_L}{(2\pi/D)^{D/2}
\, {\cal N}_L } \tow{\ =\ }{\ L \gg 1\ } L^{-D/2},\ \ \ \ {\cal N}_L =
{\rm nr.\ of\ trials\ at\  } L
\label{e.freq}
\ee
(the factor 2 appears because on the cubic lattices
we only have even loop lengths).

\section{Special configurations}

For a first test of our approach we use special gauge field configurations:
a constant magnetic field in full or half space and a constant
topological charge density.

\subsection{Homogeneous fields}

For homogeneous field configurations the continuum path integral can be
analytically evaluated to give the well known Euler-Heisenberg-Schwinger
Lagrangian (here in Euclidean space-time):
\be
{\cal L}(c,b) = \frac{1}{(4 \pi)^{D/2}} \int_0^{\infty} \frac{dT}{ T} T^{-D/2}
\e^{-m^2 T} \frac{cT}{\sinh(cT)} \frac{bT}{\sinh(bT)}
\label{e.hfc1}
\ee
where
\bear
c^2 &=& \frac{1}{2}\left[ {\vec E}^2 + {\vec B}^2 -
\sqrt{({\vec E}^2 + {\vec B}^2)^2 -4 ({\vec E}{\vec B})^2} \right]
\nonumber \\
b^2 &=& \frac{1}{2}\left[ {\vec E}^2 + {\vec B}^2 +
\sqrt{({\vec E}^2 + {\vec B}^2)^2 -4 ({\vec E}{\vec B})^2} \right]
\label{e.hfc2}
\ear
Of course before performing the $T$ integration one has to renormalize,
i.e. to subtract the integrand at $T=0$ for $D=3$ and to subtract also
its $T^2$ dependence for $D=4$
\be
\label{9}
\frac{x}{\sinh x}\to \frac{x}{\sinh x}-1-\frac{1}{6}x^2
\ee

\no The constant magnetic field can be realized on the lattice as
\bear
U_{n, {\hat 1}} &=& \exp \left(2   \pi i n_y \frac{k_y}{{\hat N}_y}\right), \ \
 n_y = 1, \dots, {\hat N}_y = N_y, \label{e.mgf11}\\
U_{n, {\hat 2}} &=& \exp \left( -2  \pi i n_x \frac{k_x}{{\hat N}_x}\right),\ \
n_x = 1, \dots, {\hat N}_x = N_x,
\label{e.mgf1}
\ear
all other links being 1. This provides a field
\be
b_x= 2 \pi k_x /{\hat N}_x, \ \  b_y = 2 \pi k_y /{\hat N}_y,\ \
b = b_x+b_y,
\label{e.mgf2}
\ee
where $k_x, k_y$ are integers, as required by the
periodic boundary conditions.

\begin{figure}[htb]
\vspace{7cm}\includegraphics{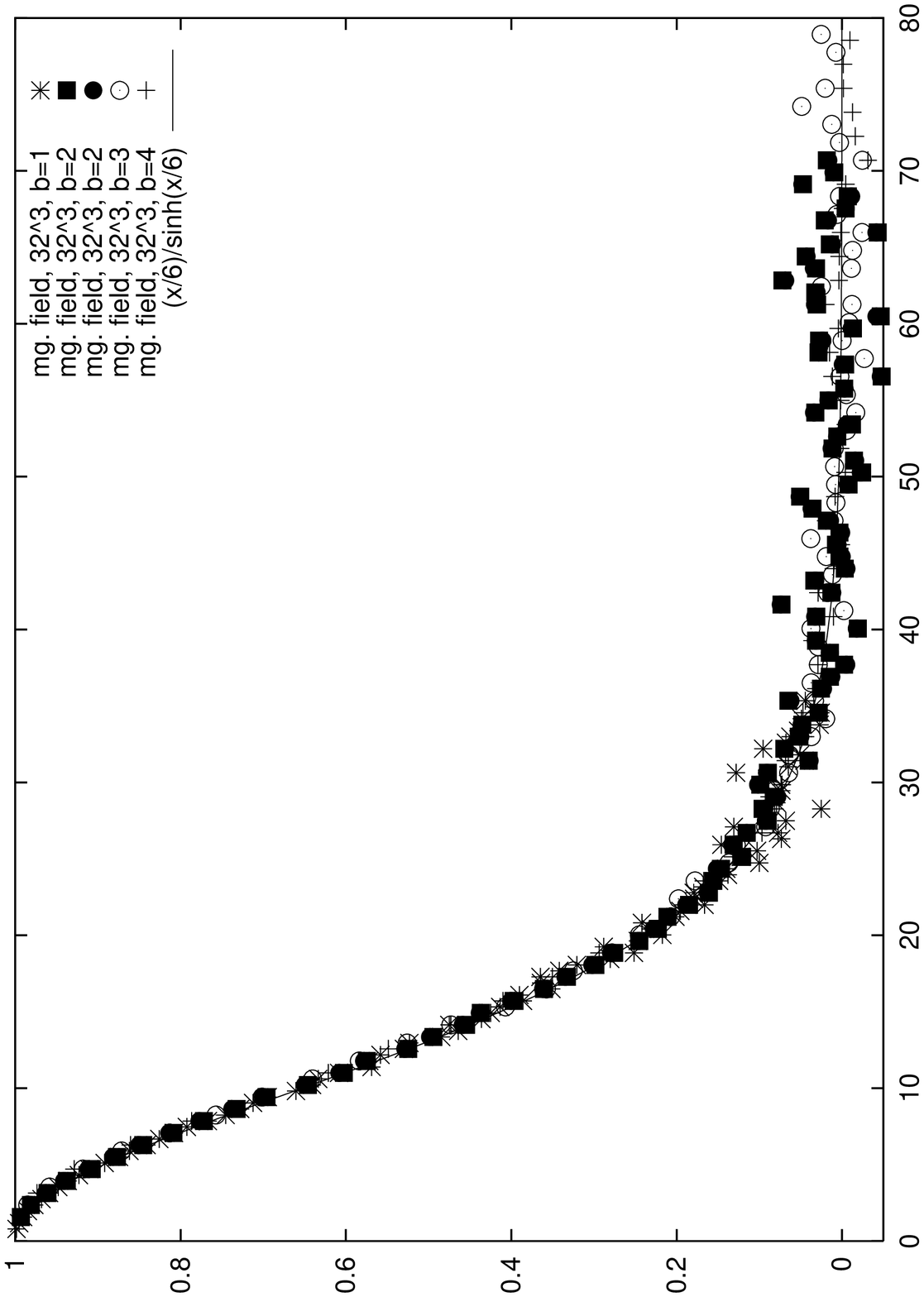}
\includegraphics{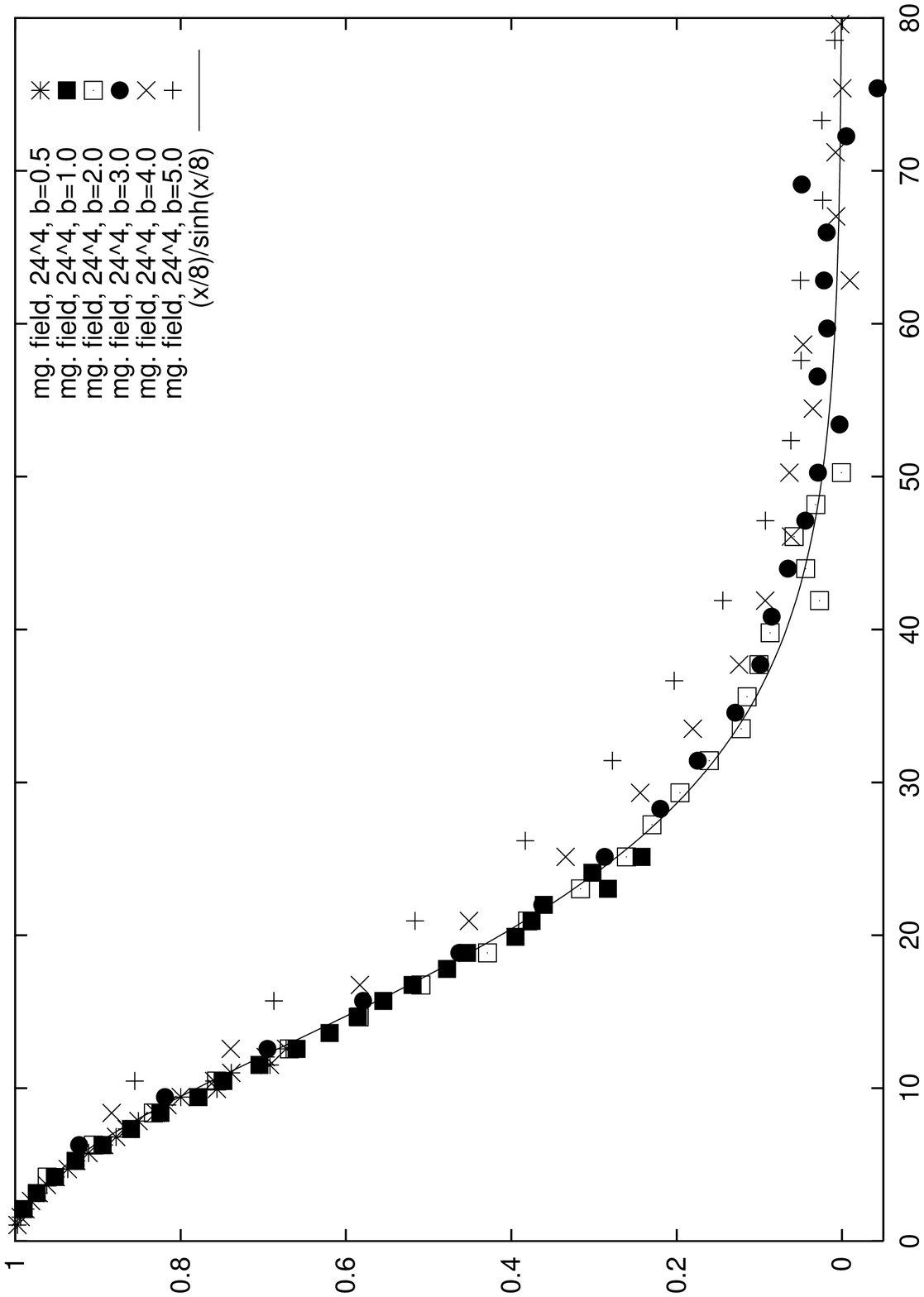}
\caption{{\it Contributions of loops of given length, $F(L,b)$ 
vs $x \equiv bL$,
in a homogeneous
magnetic field $b$   for  various $b$ in units of $\pi/8$  
on a $32^3$ lattice in $D=3$ (left  
plot) and in units of $\pi/6$ on a $24^4$ lattice in $D=4$
(right plot).}
}
\label{f.mgf3}
\end{figure}

\begin{figure}[htb]
\vspace{7cm}\includegraphics{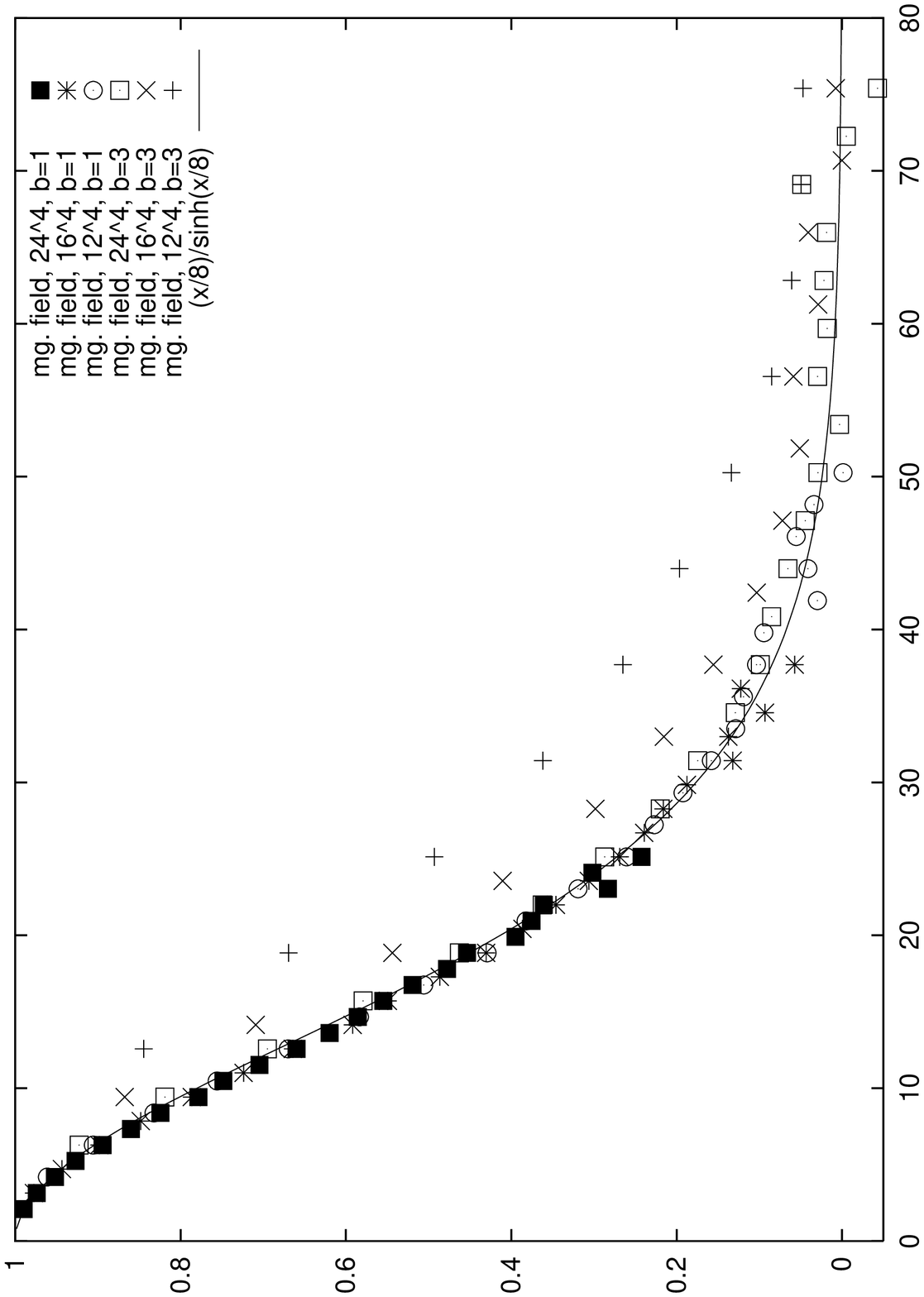}
\includegraphics{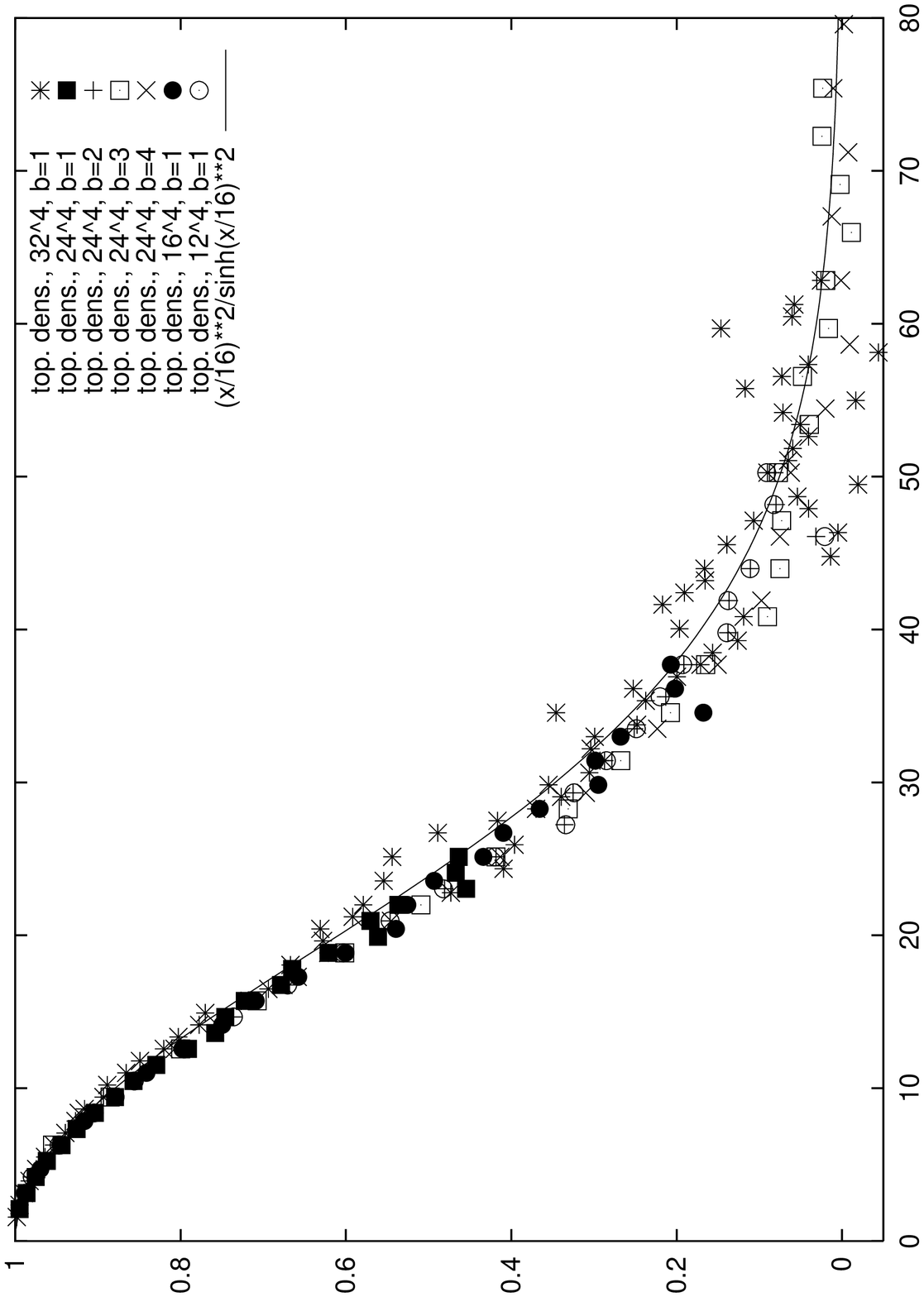}
\caption{\it Contributions of loops of given length, $F(L,b)$ vs $x \equiv bL$ 
 for various lattices  $N^4$ in $D=4$:
in a homogeneous
magnetic field $b$  ($b$ in units of $4\pi/N$)  (left  
plot) and in a  field of constant topological charge density 
$q = \frac{b^2}{8 {\pi}^2}$, $b$ in units of $2\pi / N$
(right plot).}
\label{f.tcd}
\end{figure}

In Figs. \ref{f.mgf3} and \ref{f.tcd} (left hand plot) we show the quantity:
\be
F(L,b) = \frac{1}{n_L}
\sum_{i=1}^{n_L}\Tr\left( \prod_{l \in \omega_L^i} U_l \right)
\ee
for various lattices.
From (\ref{e.hfc1},\ref{e.hfc2})  (with $B_z=b$, all
other fields zero) this should be
\be
F_c(L;b) = \frac{u}{\sinh u},\ \  u = \frac{b L}{2D}.
\label{e.fc}
\ee
For small fields the points fall well together and
reproduce the continuum result (\ref{e.fc}).
They also
scale well with the lattice size - which, by (\ref{e.mgf11}-\ref{e.mgf2}),
fixes the roughness of the potential, hence indirectly the
lattice spacing. For
large fields deviations are seen on small lattices, which
should be explained by lattice artifacts and signal the departure from
the continuum. Notice that the data in the figures are plotted against $bL$,
therefore for large $b$ we see the small $L$ contributions.

We also consider SU(2) configurations of the form:
\bear
U_{n, {\hat 1}} &=& \exp \left( 2 \pi i \sigma_3  
n_y \frac{k_y}{N_y}\right),\ \ 
n_y = 1, \dots, N_y , \label{e.to11}\\
U_{n, {\hat 3}} &=& \exp 
\left( -2  \pi i \sigma_3 n_t \frac{k_t}{N_t}\right),\ \  
n_t = 1, \dots,  N_t ,
\label{e.to1}
\ear
all other links being 1. We take $N_\mu = N$ and
\be
k_y= k_t = k;\ \ b = 2 \pi k /N
\label{e.to2}
\ee
where $k$ is an integer. Such a configuration has
${\vec E} = {\vec B} = (0,0,b)$ and hence topological charge density
\be
q = \frac{2 k^2 }{N^2} = \frac{b^2}{8 \pi^2}
\ee
homogeneously distributed over the lattice. We
 calculate
\be
F(L,b) = \frac{1}{n_L}
\sum_{i=1}^{n_L}\Tr\left( \prod_{l \in \Gamma^L_i} U_l \right),
\label{e.tcd}
\ee
shown in Fig. \ref{f.tcd}, right hand plot.
From (\ref{e.hfc1},\ref{e.hfc2}) this should be
\be
F_c(L,b) = \frac{u^2}{\sinh^2u}, \ \  u= \frac{bL}{4D}.
\ee
Again we see good agreement and scaling both with the field and
with the lattice size.

\subsection{Magnetic field in a half space}
\label{s.mghs}

\begin{figure}[htb]
\vspace{7cm}\includegraphics{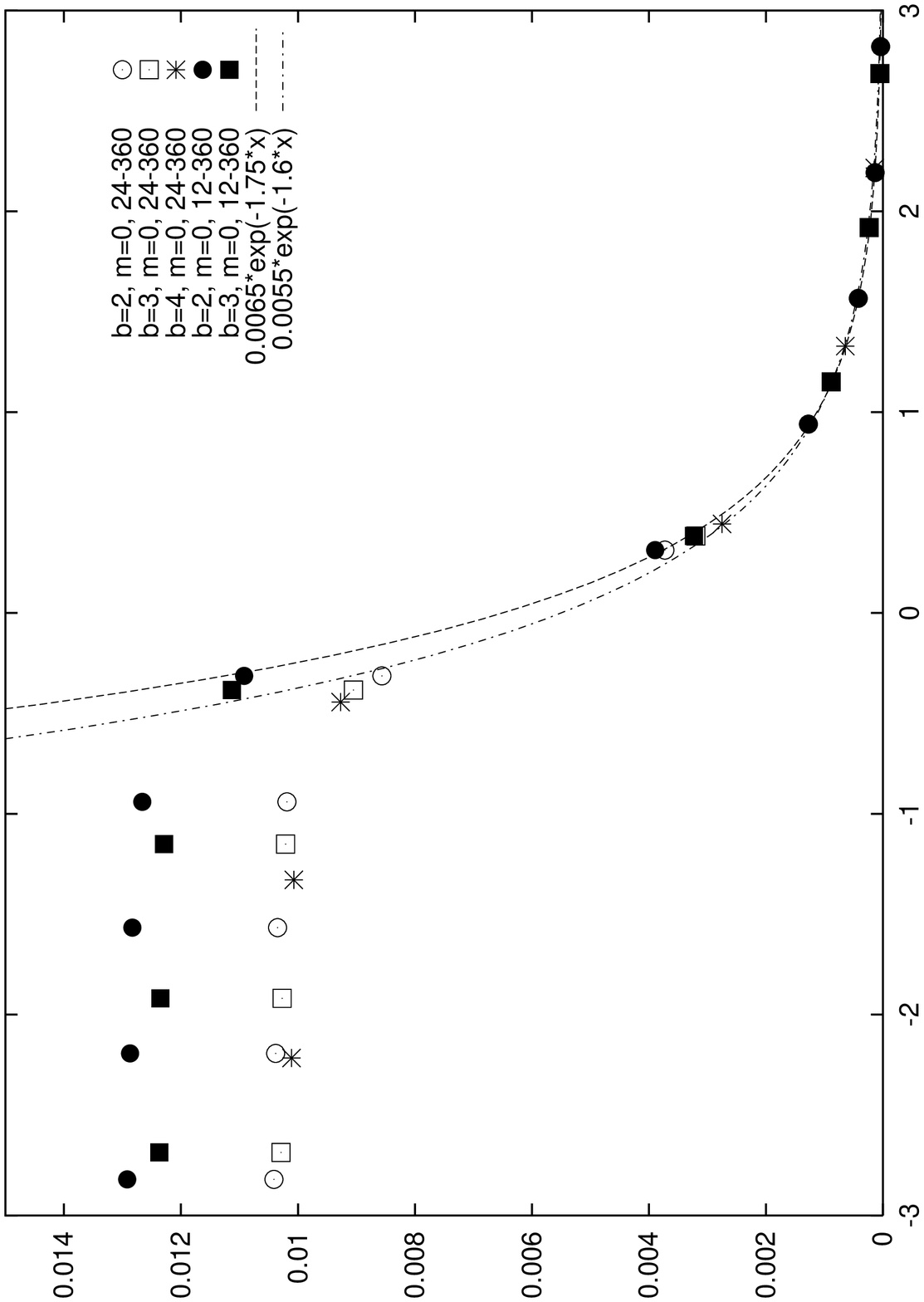}
\includegraphics{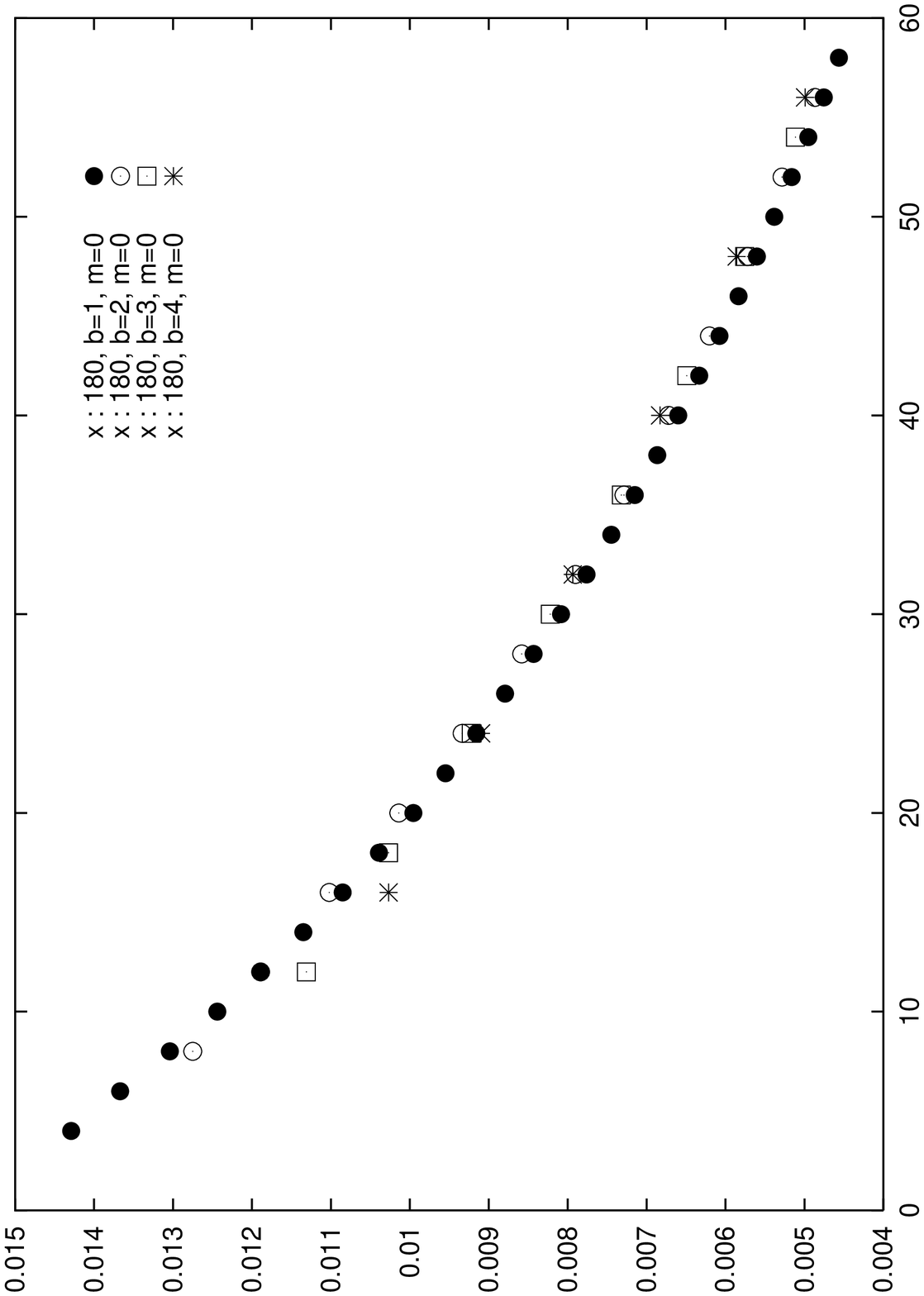}
\caption{\it Scaling propagator $f(\lambda_1,\lambda_2,\mu,\xi)$
eq. (\ref{e.scal1}) for $\mu=0$, various $b$ (in units of $\pi/16$). Left  
plot: $f$ vs $\xi$ for various summation intervals $\lambda_1\, -\, \lambda_2$.
 Right plot:
$f$ vs $\lambda_1$ for $\mu=0$,  $\lambda_2=180$ and $\xi=-5$ (non-vanishing  
field region).}
\label{f.hmb1}
\end{figure}

Let us consider now a magnetic field in $D=3$ in a half space
(x-direction) pointing into the $z$-direction.
Take ${\hat N}_x = N_x/2$ in (\ref{e.mgf1}) and $m=0$.
 We calculate
\be
{\cal L}(b,x) = \sum_L \frac{1}{L} \nu_L F(L,b,x); \ \ F(L,b,x) =
\frac{1}{n_L}
\sum_{i=1}^{n_L}\Tr\left( \prod_{l \in {\tilde \omega}_L^i(x)} U_l \right)
\label{e.G}
\ee
where $\nu_L$ are the correct loop frequencies (\ref{e.freq}). Here the loops in
${\tilde \omega}_L^i(1)$ are obtained from the original ensemble 
$\omega_L^i(1)$ by approximate centering at the point $x_{\mu}=1$ 
(and are hence confined
in a radius $\sim \sqrt{L}$ around this point). Then the 
 ${\tilde \omega}_L^i(x)$ are obtained from the ${\tilde \omega}_L^i(1)$
 by translation with
$x$ lattice units in the $x-$direction.
We denote
\be
{\cal L}(L_1,L_2,b,m,x) = \sum_{L=L_1}^{L_2} \frac{1}{L} \nu_L F(L,b,m,x)
\label{e.llu}
\ee
From the original $T-$ integral (\ref{e.wlac}), as discretized in
(\ref{e.wlal}), we should have:
\be
{\cal L}\left(\frac{\lambda_1}{b},\frac{\lambda_2}{b},b,\mu \sqrt{b},
\xi\sqrt{b}+x_0 \right) =
b^{-\frac{3}{2}} f(\lambda_1,\lambda_2,\mu,\xi).
\label{e.scal1}
\ee
where $x_0$ ($\xi = 0$) is the separation between the regions
with and without magnetic field.
We illustrate in Fig. \ref{f.hmb1}      the test of this
scaling relation. We find out that it is rather well fulfilled
if we take $x_0= \frac{N_x}{2} +0.5$ -- notice, that according to
(\ref{e.mgf1},\ref{e.mgf2}) the field extends with half strength
in the region $\frac{N_x}{2},\frac{N_x}{2}+1$ before vanishing.

\begin{figure}[htb]
\vspace{7cm}\includegraphics{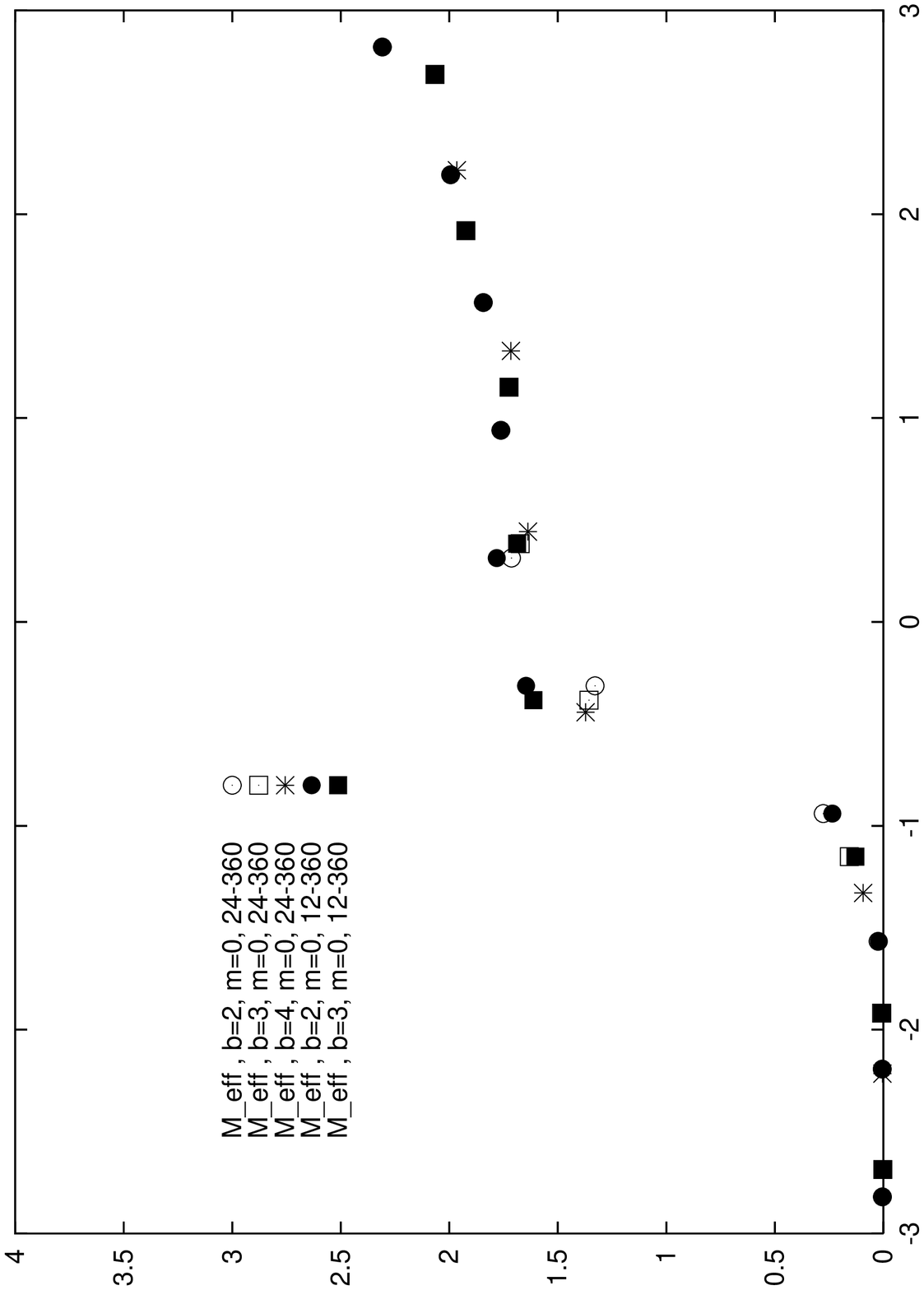}
\includegraphics{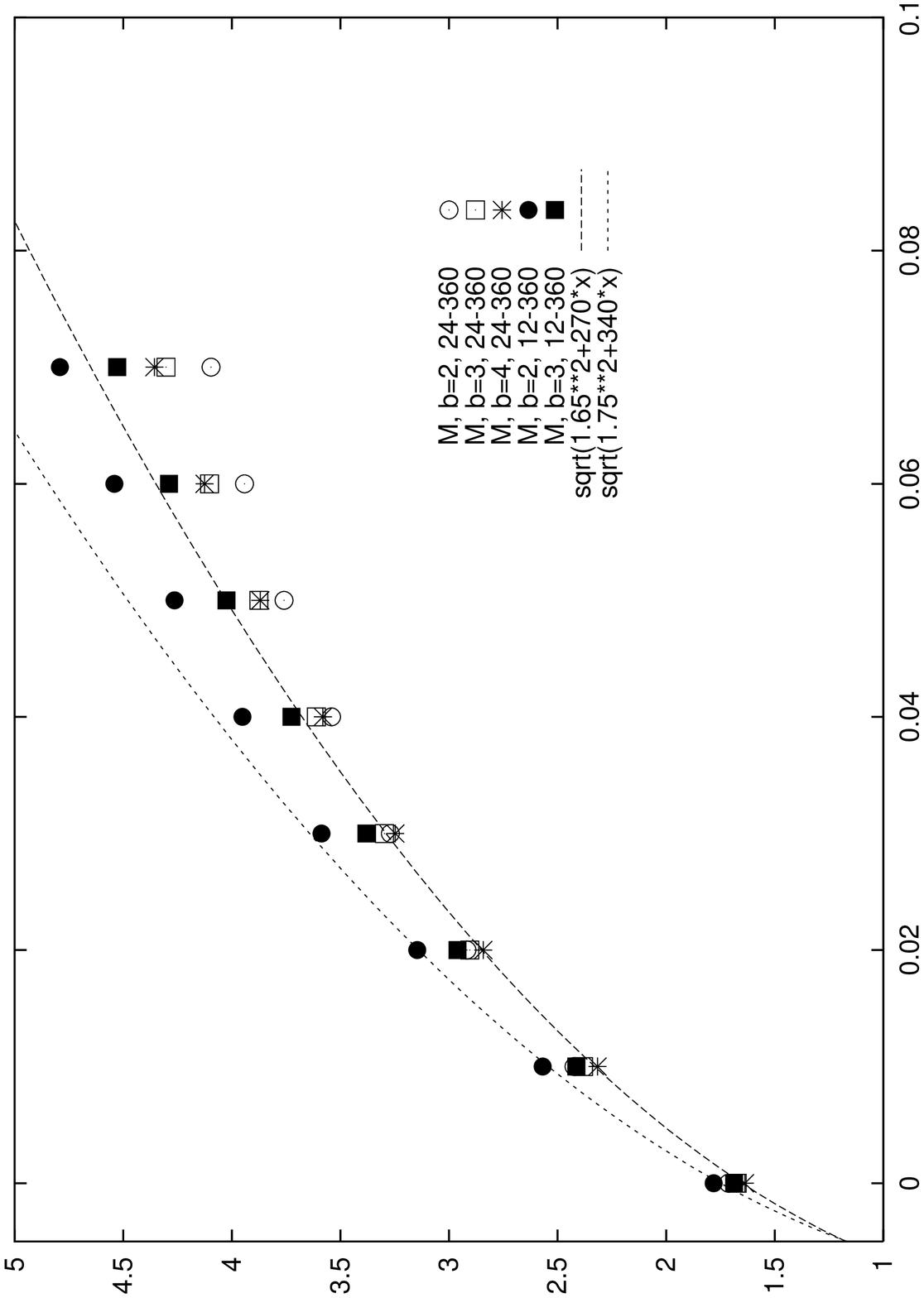}
\caption{\it Effective mass $M_{\rm eff}(\lambda_1,\lambda_2,\mu,\xi)$
eq. (\ref{e.em}) for various summation intervals $\lambda_1\, -\, \lambda_2$
and various $b$ (in units of $\pi/16$):
 vs $\xi$ for $\mu=0$ (left plot) and vs $\mu^2$, $\xi\simeq 0.5$ (right plot).
}
\label{f.hmb2}
\end{figure}

On physical grounds, see also \cite{11}, we expect
\be
f(0,\infty,\mu,\xi) \propto \e^{-M(\mu) \xi}
\label{e.exp}
\ee
In the actual calculation we have  $0 < \lambda_{1,2} < \infty$ and
discretization artifacts, which bring
systematic errors on $M$. On Fig. \ref{f.hmb1}, right hand plot, 
we illustrate the dependence of $f$ on $\lambda_1$;
$M$ itself, however, appears less sensitive -- see Fig. \ref{f.hmb1}, 
left hand plot, and Fig. \ref{f.hmb2}. In Fig. \ref{f.hmb2}    we show the ``effective mass"
\be
M_{\rm eff}(\lambda_1,\lambda_2,\mu,\xi) =
b^{-1/2}\,\ln\frac{{\cal L}\left(\frac{\lambda_1}{b},\frac{\lambda_2}{b},b,
\mu \sqrt{b},x \right) }
{{\cal L}\left(\frac{\lambda_1}{b},\frac{\lambda_2}{b},b,
\mu \sqrt{b},x+1 \right) }
\label{e.em}
\ee
We observe reasonably good scaling, at least for not too large $b$ and $m$
(where small loops dominate and therefore the
discretization artifacts are enhanced). Notice that the $m>0$ results
continue nicely to the $m=0$ ones, indicating that we do not need to take
extrapolations but can directly work at $m=0$. We read off from these
figures $M(0)=1.7 \pm 0.2$. The error is estimated from the sensitivity
to the other parameters. Since we expect statistical errors to be smaller
than this we did not attempt a jack knife analysis. Note that our result
for $M(0)$ is about a factor 2 larger than that of ref. \cite{11}. This  
disagreement is beyond the estimated error.

To understand the roots of the systematic error we observe that in
order to perform the $T$ integral, we need reliable
estimates over the full $T$ range. Consider the simpler
case of a constant field in the full space:
For small $T$ below $T=\frac{2}{D}$ (or, respectively,
$T=\frac{3}{2D}$ in the spirit of a Riemann sum) corresponding
to our smallest path with $L=4$ the integrand factor $(bT/\sinh bT
-1)$ can be safely approximated by $-\frac{1}{6}(bT)^2$
for $bT<1$. For large paths with $L\geq L_{Max}=10D/b$ corresponding
to $bT\geq5$ the $bT/\sinh bT$ can be neglected against
-1 in the integrand, i.e. we only need paths in the outer field
below $L_{Max}$. The integrand above $T_{Max}=L_{Max}/2D$ is easily
evaluated (as well as the corresponding $L$ sum producing a
$\zeta$ function). Both estimates indicate that the integrand
at small and large $L$ can be approximated by extrapolation.

\section{Discussion and Conclusions}

We have introduced the Random Walk Worldline method as a statistical
summation of the loop expansions for (logarithms of) determinants,
and have given
examples for effective
actions of rather simple gauge fields. A more exciting application would be to
lattice discretized nontrivial field configurations like instantons,
sphalerons, bounce solutions.

We had a scalar particle in the loop but it is very simple to change
to fermions and gauge bosons, just adding a spin term in the worldline
Lagrangian containing the outer field \cite{3}. In the case of
a (nonabelian) gauge field background spin $1/2$ corresponds to a
term $i\sigma_{\mu\nu}{\bf F}_{\mu\nu}$, spin 1 to $2i\ {\bf F}_{\mu\nu}$
(Feynman gauge). These terms add knots to the Wilson loop and can be
discretized as $U$-plaquettes $U_\mu U_\nu U^+_\mu U^+_\nu$ to be
joined to the path $\omega_L$ in all possible ways with a $\sigma_{\mu\nu}$ or
2-factor at the conjunction for spin $1/2$ or spin 1, respectively.
E.g. the Euler-Heisenberg formula for fermions is thus obtained from
the bosonic case very easily.
There is an overall Dirac and color trace for fermions and a
Lorentz and color trace for gauge bosons in the loop. The latter of
course require $U$-connections in the adjoint representation. The spin term
can be further modified by introducing Grassmann variables. This is very
useful in analytical calculations, in particular if worldline
supersymmetry  is used \cite{14}, \cite{3}. Also scalar/pseudoscalar
and axial vector couplings to spin $1/2$ fermions can be treated
that way. 

Alternatively we can stay with the direct expansion of the fermionic 
(logarithm of the) determinant and perform the Dirac traces the
way  it is done for the hopping parameter expansion \cite{stadet}. 

Even more interesting is the application of the method in the case
of a fluctuating gauge field background in 3- and 4-dimensional QCD and
3-dimensional thermal electroweak theory. This leads to genuine
nonperturbative phenomena.

In the general case the quenched matter free energy is
given by (\ref{e.qcd1})
where we use the ensemble of loops generated by random walk.
Varying $\beta$ we tune the lattice spacing. The size of the
lattice should be chosen such as to accommodate the loops which
contribute significantly to (\ref{e.qcd1}).

Of course one can discuss strongly interacting gauge theories
like QCD fully with lattice methods.
Still it is a relevant question how
the elementary fermion/gauge field propagators (here in a loop)
are deformed in
presence of such backgrounds. The answer might allow a semianalytical
treatment of some aspects.

In this context the discussion of chiral symmetry breaking seems to be
most promising. For our simple case of a constant $B$-field this is
easy to explain in an analytical treatment: calculate e.g.
in $d=3$ with the Banks-Casher formula for the fermion in the
loop
\bear\label{x}
<\bar\psi\psi>&=&-\frac{d}{dm}\Gamma^{(1)}(B)\Big|_{m^2=0}=-2(2m)\int^
\infty_0\frac{dT}{T}\frac{1}{(4\pi T)^{3/2}}\frac{BT}{\tanh BT}e^{-
m^2T}\Big|_{m=0}\nonumber\\
&=&-\frac{8mB}{(4\pi)^{3/2}m}\int^\infty_0dx e^{-x^2}=\frac{B}{2\pi}\ear
(this is well known \cite{16}). A discretization in case of
a fluctuating gauge field background appears to be very promising
\cite{17}. One also may like to develop some modified perturbation
theory with an IR cutoff like in the stochastic vacuum model \cite{18}
and thus discuss a ``magnetic gauge boson mass'' \cite{8},
perhaps with a tachyonic component, in this context.

\section*{Acknowledgment:} We should like to thank
Martin Reuter and Werner Wetzel for discussions. 
The calculations have been done on the VPP300
Computer at the University of Karlsruhe.

\end{document}